\def\year{2019}\relax
\documentclass[letterpaper]{article} 
\usepackage{aaai19}  
\usepackage{times}  
\usepackage{helvet}  
\usepackage{courier}  
\usepackage{url}  
\urldef{\mailsa}\path|{jiliu, leizhang}@cqu.edu.cn|
\usepackage{graphicx}  
\usepackage{amssymb}
\usepackage{amsmath}
\usepackage{algorithm}
\usepackage{algorithmic}
\usepackage{epstopdf}

\begin{document}
%
\title{Optimal Projection Guided Transfer Hashing for Image Retrieval}
\author{Ji Liu and Lei Zhang*\\
Learning Intelligence \& Vision Essential Group\\
School of Microelectronics and Communication Engineering, Chongqing University, China\\
\url{jiliu@cqu.edu.cn}, \url{leizhang@cqu.edu.cn}
}
\maketitle
\begin{abstract}
Recently, learning to hash has been widely studied for image retrieval thanks to the computation and storage efficiency of binary codes. For most existing learning to hash methods, sufficient training images are required and used to learn precise hashing codes. However, in some real-world applications, there are not always sufficient training images in the domain of interest. In addition, some existing supervised approaches need a amount of labeled data, which is an expensive process in terms of time, labor and human expertise. To handle such problems, inspired by transfer learning, we propose a simple yet effective unsupervised hashing method named Optimal Projection Guided Transfer Hashing (GTH) where we borrow the images of other different but related domain i.e., source domain to help learn precise hashing codes for the domain of interest i.e., target domain. Besides, we propose to seek for the maximum likelihood estimation (MLE) solution of the hashing functions of target and source domains due to the domain gap. Furthermore, an alternating optimization method is adopted to obtain the two projections of target and source domains such that the domain hashing disparity is reduced gradually. Extensive experiments on various benchmark databases verify that our method outperforms many state-of-the-art learning to hash methods. The implementation details are available at \url{https://github.com/liuji93/GTH}.
\end{abstract}

\section{Introduction}
In recent years, learning to hash algorithms have been proposed to handle the large-scale information retrieval problems in machine learning, computer vision,
and big data communities~\cite{Wang2017A}. The main goal of hashing techniques is to encode documents, images, and videos to a set of compact binary codes that preserve the feature similarity/dissimilarity in Hamming space. As a result, there will be less storage cost and faster computational speed by using binary features.

However, most existing learning to hash methods are faced with two problems. On one hand, most existing learning to hash methods usually require a large amount of data instances to learn a set of binary hashing codes. However, in some real-world applications, for a domain of interest, i.e., the target domain, the data instances may not be sufficient enough to learn a precise hashing model. Some supervised methods need a large number of labeled images to learn hashing codes. It is well-known that it takes a lot of time, labor and human expertise to tag images. On the other hand, they assume that the distributions of training and testing data are similar, which may not hold in many real-world applications such as cross pose and cross camera cases, etc.
\begin{figure}
\begin{center}
\includegraphics[width=0.8\linewidth]{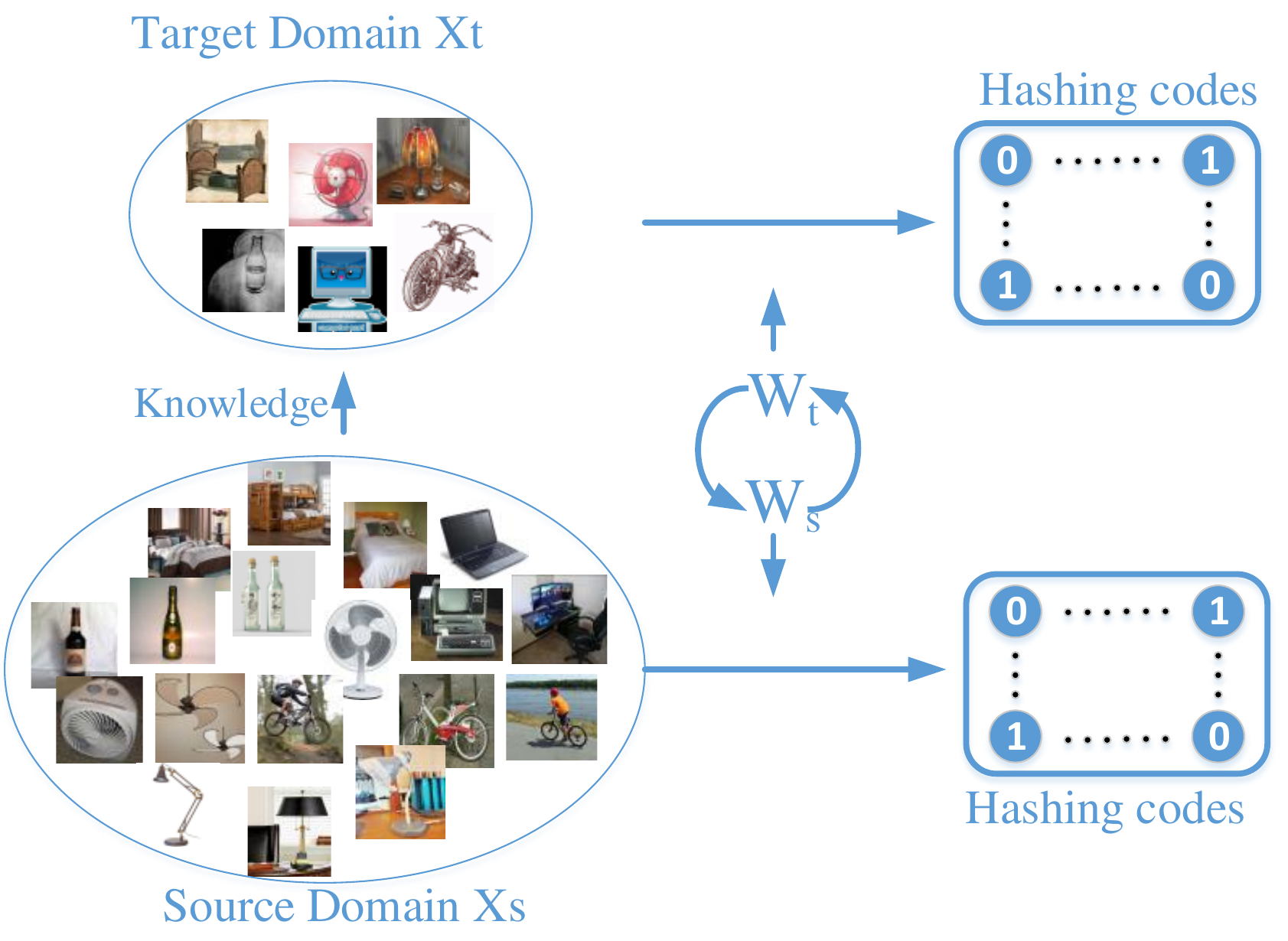}
\end{center}
   \caption{Overview of our GTH. The images from the relevant but different source domain are used to help learn hashing codes for target domain where there are insufficient images that can be used to learn effective hashing codes.}
\label{fig1}
\end{figure}

To handle the above problems, inspired by transfer learning, we propose a simple yet effective Optimal Projection Guided Transfer Hashing (GTH) method in this paper. Due to the distribution disparity of source and target domains, we propose to learn two hashing projections for target and source domains respectively in our GTH. Moreover, the knowledge from source domain can be easily used to promote target domain to learn precise hashing codes. In transfer hashing, it is important to guarantee similar images between target and source domains have similar hashing codes. In our GTH, we assume that similar images between target and source domains should mean small discrepancy between hashing projections. To this end, we let the hashing projection (functions) of target domain close to the hashing projection of source domain.

It is easy to adopt minimizing $l_2$ or $l_1$ loss between the two hashing projections of source and target domains directly. In other words, in the term of maximum likelihood estimation, we actually assume that errors between two projections of source and target domains obey Gaussian or Laplacian distribution with the $l_2$ or $l_1$ loss. However, the data distributions of source and target domains are not similar due to the existence of cross pose, cross camera, and illumination variation, etc. Therefore, the distribution of errors may be far from Gaussian or Laplacian distribution. To improve the above problem, we propose the GTH model from the view of maximum likelihood estimation in this paper. Inspired by \cite{yang2011robust}, we design
an iteratively weighted $l_2$ loss for the errors between the projections of source and target domains, which makes our GTH more adaptive to cross-domain case.

Besides, an alternating optimization method is adopted to obtain the two projections of target and source domain such that the domain disparity is reduced gradually. The two different domains can share the hashing projections each other. In other words, the target projection learning is guided by source projection and, in return, the source projection learning is guided by target projection. Finally, the optimal projections of target and source domains will be obtained. The overview of our GTH is shown as Fig. \ref{fig1}. The main contributions and novelties of this paper are summarized as follows.
\begin{itemize}
\item Guided by transfer learning, we propose a simple Optimal Projection Guided Transfer Hashing (GTH) method. To the best of our knowledge, there are few methods proposed to handle the problem that there are insufficient training images to learn precise model. We first develop a total unsupervised transfer hashing method to solve cross-domain hashing problem for image retrieval based on conventional machine learning.
\item We first propose to learn hashing projections for target and source domains respectively due to the domain disparity. The domain gap is reduced by modeling on hashing projections rather than data level.
\item In our GTH, we propose to seek for the maximum like-lihood estimation (MLE) solution of the hashing functions of target and source domains due to the domain gap, and design an iteratively weighted $l_2$ loss for the errors between the projections of source and target domains such that the high error will be punished. Besides, the projections of target and source domain are optimized in a sharing way such that the domain hashing disparity is reduced gradually.
\item Extensive experiments on various benchmark databases have been conducted. The experimental results verify that our method outperforms many state-of-the-art learning to hash methods.
\end{itemize}
\section{Related Work}
In this section, we present related works on learning to hash and transfer learning.
\subsection{Learning to hash}
In the past 10 years, various hashing methods have been proposed. Based on whether priori semantic information is used, they can categorized into two major groups: supervised hashing and unsupervised hashing. There are a lot of supervised hashing methods such as LDA hashing~\cite{Strecha2011LDAHash}, Minimal Loss Hashing~\cite{Norouzi2011Minimal}, FastHash~\cite{Lin2014Fast}, Kernel-based Supervised Hashing (KSH)~\cite{Liu2012Supervised}, Supervised Discrete Hashing (SDH)~\cite{shen2015supervised}, the Kernel-based Supervised Discrete Hashing (KSDH)~\cite{Shi2016Kernel}, and Supervised Quantization for similarity search (SQ)~\cite{Wang2016Supervised} that preserve similarity/dissimilarity of intra-class/inter-class images by using semantic information. However, there always lacks label information for model learning due to the high cost of labour and finance in some real-world application situation.

Unsupervised hashing methods aim to explore the intrinsic structure of data to preserve the similarity of neighbors without any supervised information. A number of unsupervised hashing methods have been developed in recent years. Locality-sensitive Hashing (LSH)~\cite{Gionis1999Similarity}, a typical data-independent method, uses a set of randomly generating projection to transform the image features to hashing codes. The representative unsupervised and data-dependent hashing methods include  Spectral Hashing (SH)~\cite{Weiss2008Spectral}, Anchor Graph Hashing (AGH)~\cite{Liu2011Hashing}, Iterative Quantization (ITQ)~\cite{Gong2013Iterative}, Density Sensitive Hashing (DSH)~\cite{jin2014density}, Circulant Binary Embedding (CBE)~\cite{yu2014circulant}, etc. Several ranking-preserved hashing algorithms have been proposed recently to learn more discriminative binary codes e.g., Scalable Graph Hashing (SGH)~\cite{Jiang2015Scalable}, and Ordinal Constraint Hashing (OCH)~\cite{Liu2018Ordinal}.
\subsection{Transfer learning}
Transfer learning (TL)~\cite{Pan2010A}, a new proposed learning conception, aims to transfer knowledge across two different domains such that rich source domain knowledge can be utilized to generate better classifiers on a target domain. In transfer learning, the transferred knowledge can be labels~\cite{Zhou2014Hybrid},~\cite{yang2017enhancing}, features ~\cite{Zhang2016Robust},~\cite{Xu2017Attribute},~\cite{Yang2016Zero},~\cite{Wang2017Class} and cross domain correspondences~\cite{Zhang2016LSDT},~\cite{Wang2017Adversarial}. Transfer learning has shown promising results in many machine learning tasks, such as classification and regression. To the best of our knowledge, there are few works on studying transfer learning for hashing. Most of them are based on deep learning ~\cite{Venkateswara2017Deep}. The recent work \cite{zhou2018transfer} proposes a transfer hashing from shallow to deep. Different from their works, we focus on how to transfer knowledge across hashing projection in an unsupervised manner. It is worth noting that the labels in neither of target and source domains are used in our GTH.
\section{Optimal Projection Guided Transfer Hashing}
In this section, we present the detailed discussion of Optimal Projection Guided Transfer Hashing (GTH) method.
\subsection{The objective function of our GTH}
Suppose that we have $N_t$ target data points $\mathbf{X}_t=[\mathbf{x}_{t_1}, \mathbf{x}_{t_2}, \cdots, \mathbf{x}_{t_{N_t}}] \in \mathbb{R}^{d \times{N_t}}$. We aim to learn a set of binary code $\mathbf{B}_t=\{\mathbf{b}_{t_i}\}_{i=1}^{N_t}\in\{-1,1\}^{r\times N_t}$ to well preserve feature information of the original dataset. $\mathbf{b}_{t_i}$ is the corresponding binary codes of $\mathbf{x}_{t_i}$. $N_t$, $d$, and $r$ denote the number of the target domain samples, the dimension of each sample, and the code length of binary feature, respectively. Similar with most of learning to hash methods, we also learn hashing projection to map and quantize each $\mathbf{x}_{t_i}$ into a binary codes $\mathbf{b}_{t_i}$. However, when the available target training data is limited, i.e., $N_t$ is small, the binary codes learned by existing learning to hash methods can't perform well. In our GTH, we take advantage of the knowledge (i.e., features) of another known domain (i.e., source domain). Suppose that we have already obtained $N_s$ source data points $\mathbf{X}_s=[\mathbf{x}_{s_1}, \mathbf{x}_{s_2}, \cdots, \mathbf{x}_{s_{N_s}}] \in \mathbb{R}^{d \times{N_s}}$.

We denote $\mathbf{B}_{t}={\rm{H}}(\mathbf{W}_t^{\rm T}\mathbf{X}_{t})$  and $\mathbf{B}_{s}={\rm{H}}(\mathbf{W}_s^{\rm T}\mathbf{X}_{s})$ where $\mathbf{W}_t \in \mathbb{R}^{d \times r}$ is hashing projection of target domain and $\mathbf{W}_s \in \mathbb{R}^{d \times r}$ is hashing projection of source domain. ${\rm{H}}(v)=sgn(v)$ equals to 1 if $v\ge0$ and -1 otherwise. In our GTH, to reduce the distribution discrepancy, we let hashing projection of target domain close to source domain:
\begin{equation}
\min_{\mathbf{W}_t, \mathbf{W}_s}\lVert{\mathbf{W}_t-\mathbf{W}_s\rVert^2}.
\label{eq0}
\end{equation}

We denote that $\mathbf{E}=\mathbf{W}_t-\mathbf{W}_s$ represents the error matrix. $E_{ij}$ is one element in the error matrix. As discussed above, from the view of maximum likelihood estimation (MLE), the error matrix follows Gaussian distribution by using the Eq.\ref{eq0}. However, the different data distributions of source and target domains may lead to that the probability distribution of error matrix is far from Gaussian distribution. Without loss of generality, we let $\mathbf{e}=[E_{11}, E_{21}, \cdots, E_{d1}, \cdots, E_{1r}, E_{2r}, \cdots, E_{dr}]^{\rm{T}}$. Assume that $e_1, e_2, \cdots, e_{N}$ are independently and identically distributed according to some probability density function (PDF) $f_{\mathbf{\theta}}(e_n)$ where $N=d \times r$ and $\mathbf{\theta}$ denotes the parameter set that characterizes the distribution. The likelihood estimation can be represented as $L_{\mathbf{\theta}}=\prod_{n=1}^Nf_{\mathbf{\theta}}(e_n)$ and MLE aims to maximize this likelihood function or minimize the negative log likelihood function: $-\rm{ln}L_{\mathbf{\theta}}=\sum_{n=1}^N{\rho_{\mathbf{\theta}}(e_n)}$ where $\rho_{\mathbf{\theta}}(e_n)=-\rm{ln}f_{\mathbf{\theta}}(e_n)$.

With the above analysis, the Eq. \ref{eq0} with uncertain probability density function can be transformed into the following minimization problem:
\begin{equation}
\min_{\mathbf{W}_t, \mathbf{W}_s}\sum_{n=1}^N\rho_{\mathbf{\theta}}(e_n).
\label{eq1}
\end{equation}

In general, we assume that the unknown PDF $f_{\mathbf{\theta}}(e_n)$ is symmetric, and the bigger error will assign a low probability value $f_{\mathbf{\theta}}(e_i)< f_{\mathbf{\theta}}(e_j)$ if $|e_i|> |e_j|$. Therefore, $\rho_{\mathbf{\theta}}(e_n)$ has the following properties: $\rho_{\mathbf{\theta}}(0)$ is the global minimal of $\rho_{\mathbf{\theta}}(e_n)$. Specially, we denote $\rho_{\mathbf{\theta}}(0)=0$; $\rho_{\mathbf{\theta}}(e_n)=\rho_{\mathbf{\theta}}(-e_n)$; $\rho_{\mathbf{\theta}}(e_i)< \rho_{\mathbf{\theta}}(e_j)$ if $|e_i|<|e_j|$.

Denote that $F_\mathbf{\theta}(\mathbf{e})=\sum_{n=1}^N\rho_{\mathbf{\theta}}(e_n)$. We approximate $F_\mathbf{\theta}(\mathbf{e})$ by using its first order Taylor expansion in the neighborhood $\mathbf{e}_0$:
\begin{equation}
\widetilde{F_\mathbf{\theta}}(\mathbf{e})=F_\mathbf{\theta}(\mathbf{e}_0)+(\mathbf{e}-\mathbf{e}_0)^{\rm{T}}F'_\mathbf{\theta}(\mathbf{e}_0)+R_1(\mathbf{e}),
\label{eq2}
\end{equation}
where $R_1(\mathbf{e})$ is the second-order remained term, and $F'_\mathbf{\theta}(\mathbf{e}_0)$ is the derivative of $F_\mathbf{\theta}(\mathbf{e}_0)$.
\begin{equation}
R_1(\mathbf{e})=0.5(\mathbf{e}-\mathbf{e}_0)^{\rm{T}}\mathbf{\Omega}(\mathbf{e}-\mathbf{e}_0).
\label{eq3}
\end{equation}
$\mathbf{\Omega}$ is a diagonal matrix and we denote
\begin{equation}
\Omega_{nn}={\rho'_{\mathbf{\theta}}(\Lambda_{n})}/{\Lambda_{n}}=\omega_{\theta}(\Lambda_{n}),
\label{eq4}
\end{equation}
where we randomly assign a value to $\Lambda_{n}$ which satisfies  $\Lambda_{n}\in(0,e_n)$ if $e_n>0$ otherwise $\Lambda_{n}\in(e_n,0)$. $\rho'_{\mathbf{\theta}}(\Lambda_{n})$ represents first derivative. Because $\rho_{\mathbf{\theta}}(0)$ is the global minimal of $\rho_{\mathbf{\theta}}(e_n)$, we can get $\rho'_{\mathbf{\theta}}(0)=0$.
We denote $\mathbf{e}_0=\mathbf{0}$ such that we can obtain the following objective function
\begin{equation}
\widetilde{F_\mathbf{\theta}}(\mathbf{e})=R_1(\mathbf{e})=0.5\lVert\mathbf{\Omega}^{\frac{1}{2}}\mathbf{e}\rVert^2.
\label{eq5}
\end{equation}
It is obvious that each element $\Omega_{nn}$ in the diagonal matrix $\mathbf{\Omega}$ can be regarded as a weight coefficient to each error value $e_n$. We expect that the higher value $|e_n|$ will be assigned a lower weight coefficient $\Omega_{nn}$.

According to \cite{yang2011robust} and \cite{Zhang2003Modified}, we also choose the signmoid function as the weight function
\begin{equation}
\omega_{\theta}(\Lambda_{n})=\rm{exp}(\mu \delta-\mu \Lambda_{n}^2)/(1+\rm{exp}(\mu \delta-\mu \Lambda_{n}^2)),
\label{eq6}
\end{equation}
where $\mu$ and $\delta$ are positive scalars. Parameter $\mu$ controls the decreasing rate from 1 to 0, and $\delta$ controls the location of demarcation point. For the choice of $\mu$ and $\delta$, we just follow \cite{yang2011robust}. Considering the
Eq.\ref{eq4}, Eq.\ref{eq6}, and $\rho_{\mathbf{\theta}}(0)=0$, we obtain $\rho_{\mathbf{\theta}}(\Lambda_{n})$ as following
\begin{equation}
\rho_{\mathbf{\theta}}(\Lambda_{n})=\frac{-1}{2\mu}(ln(1+\rm{exp}(\mu \delta-\mu \Lambda_{n}^2)-\rm{ln}(1+exp(\mu\delta))).
\label{eq7}
\end{equation}

Therefore, we can transform Eq.\ref{eq5} into matrix form as following objective function.
\begin{equation}
\min_{\mathbf{W}_t, \mathbf{W}_s}\frac{1}{2}\lVert{\mathbf{M}^{\frac{1}{2}}\odot(\mathbf{W}_t-\mathbf{W}_s)\rVert^2}.
\label{eq8}
\end{equation}
We denote $M_{ij}=\omega_{\theta}(\widetilde{E_{ij}})$ where we randomly choose a value as $\widetilde{E_{ij}}$ which satisfies $\widetilde{E_{ij}}\in(0,E_{ij})$ if $E_{ij}>0$ otherwise $\widetilde{E_{ij}}\in(E_{ij},0)$. Note that $\mathbf{M}$ is the matrix form of all diagonal elements in $\mathbf{\Omega}$.

It is worth noting that the Eq.\ref{eq8} can be viewed as a inductive model. If we let $\omega_{\theta}(\widetilde{E_{ij}})=2$, the Eq.\ref{eq8} is just Eq.\ref{eq0} which assumes that the errors obey Gaussian distribution. Specially, in this paper, GTH-h refers to Eq.\ref{eq8} with $\omega_{\theta}(\widetilde{E_{ij}})$ being Eq.\ref{eq6} and GTH-g refers to Eq.\ref{eq8} with $\omega_{\theta}(\widetilde{E_{ij}})=2$.

The quantization loss between hashing codes and its magnitude is used as regularization term in GTH. Besides, we impose orthogonality constraints to hashing projections. The overall objective function is as following
\begin{equation}
\begin{split}
& \hspace*{1.3cm}\min_{\mathbf{W}_t, \mathbf{W}_s, \mathbf{B}_t, \mathbf{B}_s}\frac{1}{2}{\lVert\mathbf{M}^{\frac{1}{2}}\odot(\mathbf{W}_t-\mathbf{W}_s)\rVert^2}\\
& \hspace*{0.2cm}+\frac{\lambda_1}{2}\lVert\mathbf{B}_t-{\rm{H}}({\mathbf{W}_t}^{\rm{T}}\mathbf{X}_t)\rVert^2+\frac{\lambda_2}{2}\lVert\mathbf{B}_s-
{\rm{H}}({\mathbf{W}_s}^{\rm{T}}\mathbf{X}_s)\rVert^2\\
& \hspace*{1.5cm} s.t.\quad  {\mathbf{W}_t}^{\rm{T}}\mathbf{W}_t=\mathbf{I}, {\mathbf{W}_s}^{\rm{T}}\mathbf{W}_s=\mathbf{I},
\end{split}
\label{eq9}
\end{equation}
where $\lambda_1 $and $\lambda_2$ denote the regularization coefficients.
\subsection{Optimization}
In this paper, we propose a weighted $l_2$ loss for the errors between the projections of source and target domains, and update the weight coefficients by using the errors from the last iteration. As the non-convex $sgn(\cdot)$ function makes Eq. \ref{eq9} a NP-hard problem, we relax the $sgn(x)$ function as its signed magnitude $x$~\cite{Lazebnik2011Iterative}. Therefore, the Eq. \ref{eq9} can be rewritten as
\begin{equation}
\begin{split}
& \hspace*{1.3cm}\min_{\mathbf{W}_t, \mathbf{W}_s, \mathbf{B}_t, \mathbf{B}_s}\frac{1}{2}{\lVert\mathbf{M}^{\frac{1}{2}}\odot(\mathbf{W}_t-\mathbf{W}_s)\rVert^2}\\
& \hspace*{0.5cm}+\frac{\lambda_1}{2}\lVert\mathbf{B}_t-{\mathbf{W}_t}^{\rm{T}}\mathbf{X}_t\rVert^2+\frac{\lambda_2}{2}\lVert\mathbf{B}_s-{\mathbf{W}_s}^{\rm{T}}\mathbf{X}_s\rVert^2\\
& \hspace*{1.5cm} s.t.\quad  {\mathbf{W}_t}^{\rm{T}}\mathbf{W}_t=\mathbf{I}, {\mathbf{W}_s}^{\rm{T}}\mathbf{W}_s=\mathbf{I}.
\end{split}
\label{eq10}
\end{equation}
As mentioned above, we will adopt a relax way to solve problem (\ref{eq9}). The solutions for optimization problem (\ref{eq10}) can be calculated by alternatingly updating the variables, $\mathbf{W}_t$, $\mathbf{W}_s$, $\mathbf{B}_t$, $\mathbf{B}_s$, and $\mathbf{M}$.

\textbf{$\mathbf{W}_t$-Step.} By fixing $\mathbf{W}_s$, $\mathbf{B}_t$, $\mathbf{B}_s$, and $\mathbf{M}$, the projection of target domain $\mathbf{W}_t$ can be obtained by solving the following subproblem
\begin{equation}
\begin{split}
&\min_{\mathbf{W}_t}{\Vert{\mathbf{M}^{\frac{1}{2}}\odot(\mathbf{W}_t-\mathbf{W}_s)\Vert^2}
+\lambda_1\Vert\mathbf{B}_t-{\mathbf{W}_t}^{\rm{T}}\mathbf{X}_t\Vert^2}\\
& \hspace*{2cm} \quad s.t.\quad  {\mathbf{W}_t}^{\rm{T}}\mathbf{W}_t=\mathbf{I}.
\end{split}
\label{eq11}
\end{equation}
Updating $\mathbf{W}_t$ is a typical optimization problem with orthogonality constraints. We apply the optimization procedure in~\cite{Wen2013A} to update $\mathbf{W}_t$. Let $\mathbf{G}_t$ be the partial derivative of the objective function with respect to $\mathbf{W}_t$. $\mathbf{G}_t$ is represented as
\begin{equation}
\mathbf{G}_t=\mathbf{M}\odot(\mathbf{W}_t-\mathbf{W}_s)+\lambda_1(\mathbf{X}_t\mathbf{X}_t^{\rm{T}}\mathbf{W}_t-\mathbf{X}_t\mathbf{B}_t^{\rm{T}}).
\label{eq12}
\end{equation}
To preserve the orthogonality constraint on $\mathbf{W}_t$, we first define the skew-symmetric matrix $\mathbf{Q}_t$~\cite{Armstrong2005Numerical} as
$\mathbf{Q}_t=\mathbf{W}_t^{\rm{T}}\mathbf{G}_t-\mathbf{G}_t^{\rm{T}}\mathbf{W}_t$.
Then, we adopt Crank Nicolson like scheme~\cite{Wen2013A} to update the orthogonal matrix $\mathbf{W}_t$:
\begin{equation}
\mathbf{W}_t^{(k+1)}=\mathbf{W}_t^{(k)}-\frac{\tau}{2}(\mathbf{W}_t^{(k+1)}+\mathbf{W}_t^{(k)})\mathbf{Q}_t,
\label{eq13}
\end{equation}
where $\tau$ denotes the step size. We empirically set $\tau=0.1$. By solving Eq. \ref{eq13}, we can get
\begin{equation}
\mathbf{W}_t^{(k+1)}=\mathbf{W}_t^{(k)}\mathbf{Q}_t,
\label{eq14}
\end{equation}
and $\mathbf{Q}_t^{(k+1)}=(\mathbf{I}+\frac{\tau}{2}\mathbf{Q}_t)^{-1}(\mathbf{I}-\frac{\tau}{2}\mathbf{Q}_t)$.
We iteratively update $\mathbf{W}_t$ several times based on Eq. \ref{eq14} with the Barzilai-Borwein (BB) method~\cite{Wen2013A}.

\textbf{$\mathbf{W}_s$-Step.} By fixing $\mathbf{W}_t$, $\mathbf{B}_t$, $\mathbf{B}_s$, and $\mathbf{M}$, the projection of source domain $\mathbf{W}_s$ can be solved as:
\begin{equation}
\begin{split}
&\min_{\mathbf{W}_s}{\Vert\mathbf{M}^{\frac{1}{2}}\odot(\mathbf{W}_t-\mathbf{W}_s)\Vert^2}
+\lambda_2\Vert\mathbf{B}_s-{\mathbf{W}_s}^{\rm{T}}\mathbf{X}_s\Vert^2\\
& \hspace*{2cm}\quad s.t.\quad  {\mathbf{W}_s}^{\rm{T}}\mathbf{W}_s=\mathbf{I}.
\end{split}
\label{eq15}
\end{equation}
Updating $\mathbf{W}_s$ is the same as $\mathbf{W}_t$. Let $\mathbf{G}_s$ be the partial derivative of the objective function with respect to $\mathbf{W}_s$. $\mathbf{G}_s$ is represented as
\begin{equation}
\mathbf{G}_s=\mathbf{M}\odot(\mathbf{W}_s-\mathbf{W}_t)+\lambda_2(\mathbf{X}_s\mathbf{X}_s^{\rm{T}}\mathbf{W}_s-\mathbf{X}_s\mathbf{B}_s^{\rm{T}}).
\label{eq16}
\end{equation}
\begin{algorithm}
\caption{Optimal Projection Guided Transfer Hashing}
\label{Algorithm1}
\begin{algorithmic}[1]
\REQUIRE Target samples $\mathbf{X}_t$ and source samples $\mathbf{X}_s$ parameters $\lambda_1=0.1$, and $\lambda_2=1$, identity matrix $\mathbf{I}$
\ENSURE  $\mathbf{W}_t$, $\mathbf{B}_t$, $\mathbf{W}_s$, and $\mathbf{B}_s$
\STATE \textbf{Initialize:} Initialize $\mathbf{W}_t^{(0)}$ and $\mathbf{W}_s^{(0)}$ as the top $r$ eigenvectors of $\mathbf{X}_t\mathbf{X}_t^{\rm{T}}$ and $\mathbf{X}_s\mathbf{X}_s^{\rm{T}}$ corresponding to the $r$ largest eigenvalues, respectively. $\mathbf{B}_t^{(0)}$ and $\mathbf{B}_s^{(0)}$ are random matrices. $k=1$.\\
\REPEAT
\STATE update $\mathbf{M}^{(k)}$: by solving $\omega_{\theta}(\mathbf{W}_t^{(k-1)}-\mathbf{W}_s^{(k-1)})$;
\STATE update $\mathbf{W}_t^{(k)}$: by solving Eq. \ref{eq14};
\STATE update $\mathbf{W}_s^{(k)}$: by solving Eq. \ref{eq18};
\STATE update $\mathbf{B}_t^{(k)}$: by solving Eq. \ref{eq19};
\STATE update $\mathbf{B}_s^{(k)}$: by solving Eq. \ref{eq20};
\STATE k=k+1;
\UNTIL {max iterations}
\end{algorithmic}
\end{algorithm}
To preserve the orthogonality constraint on $\mathbf{W}_s$, we define the skew-symmetric matrix $\mathbf{Q}_s$ as
$\mathbf{Q}_s=\mathbf{W}_s^{\rm{T}}\mathbf{G}_s-\mathbf{G}_s^{\rm{T}}\mathbf{W}_s$.
Then, we adopt Crank Nicolson like scheme to update the orthogonal matrix $\mathbf{W}_s$:
\begin{equation}
\mathbf{W}_s^{(k+1)}=\mathbf{W}_s^{(k)}-\frac{\tau}{2}(\mathbf{W}_s^{(k+1)}+\mathbf{W}_s^{(k)})\mathbf{Q}_s,
\label{eq17}
\end{equation}
where $\tau$ denotes the step size. We empirically set $\tau=0.1$ same as updating $\mathbf{W}_t$. By solving Eq. \ref{eq17}, we can get
\begin{equation}
\mathbf{W}_s^{(k+1)}=\mathbf{W}_s^{(k)}\mathbf{Q}_s,
\label{eq18}
\end{equation}
and $\mathbf{Q}_s^{(k+1)}=(\mathbf{I}+\frac{\tau}{2}\mathbf{Q}_s)^{-1}(\mathbf{I}-\frac{\tau}{2}\mathbf{Q}_s)$.
We iteratively update $\mathbf{W}_s$ several times based on Eq. \ref{eq18} with the Barzilai-Borwein (BB) method.

\textbf{$\mathbf{B}_t$-Step and $\mathbf{B}_s$-Step .} As $\mathbf{B}_t$ and $\mathbf{B}_s$ are two binary matrixes, the solutions can be directly obtained as:
\begin{equation}
\mathbf{B}_t=sgn(\mathbf{W}_t^{\rm{T}}\mathbf{X}_t).
\label{eq19}
\end{equation}
\begin{equation}
\mathbf{B}_s=sgn(\mathbf{W}_s^{\rm{T}}\mathbf{X}_s).
\label{eq20}
\end{equation}

\textbf{$\mathbf{M}$-Step.} The weight matrix $\mathbf{M}$ is directly computed as following:
\begin{equation}
\mathbf{M}=\omega_{\theta}(\mathbf{W}_t-\mathbf{W}_s).
\label{eq21}
\end{equation}
The overall solving procedures are summarized in Algorithm \ref{Algorithm1}.

\section{Experiment}
In this section, extensive experiments are conducted to evaluate the proposed hashing method on image retrieval performance. We perform the experiments on three groups benchmark datasets: PIE-C29\&PIE-C05 from \textbf{PIE}~\cite{Sim2002The}, Amazon\&Dslr from \textbf{Office}~\cite{Saenko2010Adapting}, and VOC2007\&Caltech101 from \textbf{VLCS}~\cite{Torralba2011Unbiased}. We also choose five state-of-the-art learning-to-hash methods, LSH~\cite{Gionis1999Similarity}, ITQ~\cite{Gong2013Iterative}, CBE~\cite{yu2014circulant}, DSH~\cite{jin2014density}, and OCH~\cite{Liu2018Ordinal} as baselines. For fair comparison, we introduce a NoDA method acted as OCH method.
\subsection{Datasets, Settings, and Retrieval evaluation}
\textbf{Description of Datasets:} The \textbf{PIE} dataset consists of 41,368 face images from 68 subjects as a whole. The images are under five near frontal poses (C05, C07, C09, C27 and C29). We use two subsets chosen from poses C05 and C29. Each image is resized to $32 \times 32$ and represented by a 1024-dim vector. We use pose C29 (containing 1632 images) as target domain and pose C05 (containing 3332 images) as source domain. Specially, for target domain, we randomly select 500 samples as testing images and the rest samples as training images.
\begin{table*}
\caption{The MAP scores (\%) on PIE, Amazon\&Dslr, and VOC2007\&Caltech101 databases with varying code length from 16 to 64.}
\resizebox{\textwidth}{22mm}{
\begin{tabular}{c|ccccc|ccccc|ccccc}
\hline
        &  \multicolumn{5}{c|}{PIE-C29\&PIE-C05}  & \multicolumn{5}{c|}{Amazon\&Dslr}      & \multicolumn{5}{c}{VOC2007\&Caltech101}\\
\hline
 Bit    &    16      & 24   &  32  &  48    &  64    &16    &24     &32      &48     &64     &16     &24     &32     &48     &64        \\
\hline
 LSH    &   18.23    &21.79 & 25.26& 29.91  & 32.96  &19.69 &28.92  &35.12   &46.72  &53.07  &11.06  &16.51  &20.61  &27.41  &33.12     \\
 ITQ    &   18.17    &21.63 & 23.74& 26.82  & 28.86  &43.15 &51.74  &56.80   &62.47  &65.84  &21.69  &28.52  &33.46  &39.50  &42.34     \\
 CBE    &   16.31    &22.13 & 27.10& 30.06  & 32.51  &20.82 &27.60  &36.21   &47.52  &51.96  &11.04  &15.64  &20.68  &26.97  &33.84     \\
 DSH    &   17.05    &19.60 & 22.01& 25.65  & 28.12  &26.51 &32.34  &37.39   &48.29  &50.12  &8.69   &6.23   &13.40  &15.56  &20.21     \\
 OCH    &   20.75    &26.29 & 28.96& 33.33  & 34.39  &41.77 &52.41  &56.00   &62.38  &65.45  &\textbf{32.94}  &35.45  &38.00  &41.46  &42.25     \\
 NoDA   &   21.06    &24.76 & 26.51& 32.11  & 32.34  &41.64 &51.96  &57.21   &63.29  &65.63  &30.77  &34.81  &36.95  &40.78  &41.80     \\
  \hline
 GTH-g  &   24.16    &28.40 & 31.69& 34.95  & 35.70  &44.16 &\textbf{53.57}  &\textbf{57.59}   &\textbf{63.91}  &\textbf{66.96}  &28.62  &\textbf{41.20}  &46.42  &56.59  &63.10     \\
 GTH-h  & \textbf{25.45}&\textbf{29.42} &\textbf{31.76}&\textbf{35.25}&\textbf{36.56}  &\textbf{45.23} &52.36  &57.26   &63.17  &65.63  &30.05  &39.70  &\textbf{48.14}  &\textbf{57.33}  &\textbf{63.53}     \\
\hline
\end{tabular}}
\label{tab1}
\end{table*}

The \textbf{Office} dataset is a most popular benchmark dataset for object recognition in the domain adaptation computer vision community. The dataset consists of daily objects in an office environment. \textbf{Office} has 3 domains: Amazon (A), Dslr (D), and Webcam (W). We use Amazon with 2817 images as the source domain and Dslr with 498 images as target domain. 100 images from target domain are randomly selected as testing images and the rest images are used as training images. Each image is represented by a 4096-d CNN feature vector \cite{Donahue2013DeCAF}.

The \textbf{VLCS} aggregates photos from Caltech, LabelMe, Pascal VOC 2007 and SUN09. It provides a 5-way multiclass benchmark on the five common classes: ¡¯bird¡¯, ¡¯car¡¯, ¡¯chair¡¯, ¡¯dog¡¯ and ¡¯person¡¯. The VOC 2007 dataset containing 3376 images is used as source domain and Caltech containing 1415 images is used as target domain. 100 images from target domain are randomly selected as testing images and the rest images are used as training images. Each image is represented by a 4096-d CNN feature vector \cite{Donahue2013DeCAF}.
\begin{figure*}
\centering
\begin{minipage}{5.5cm}
 \centerline{\includegraphics[height=6cm,width=6cm]{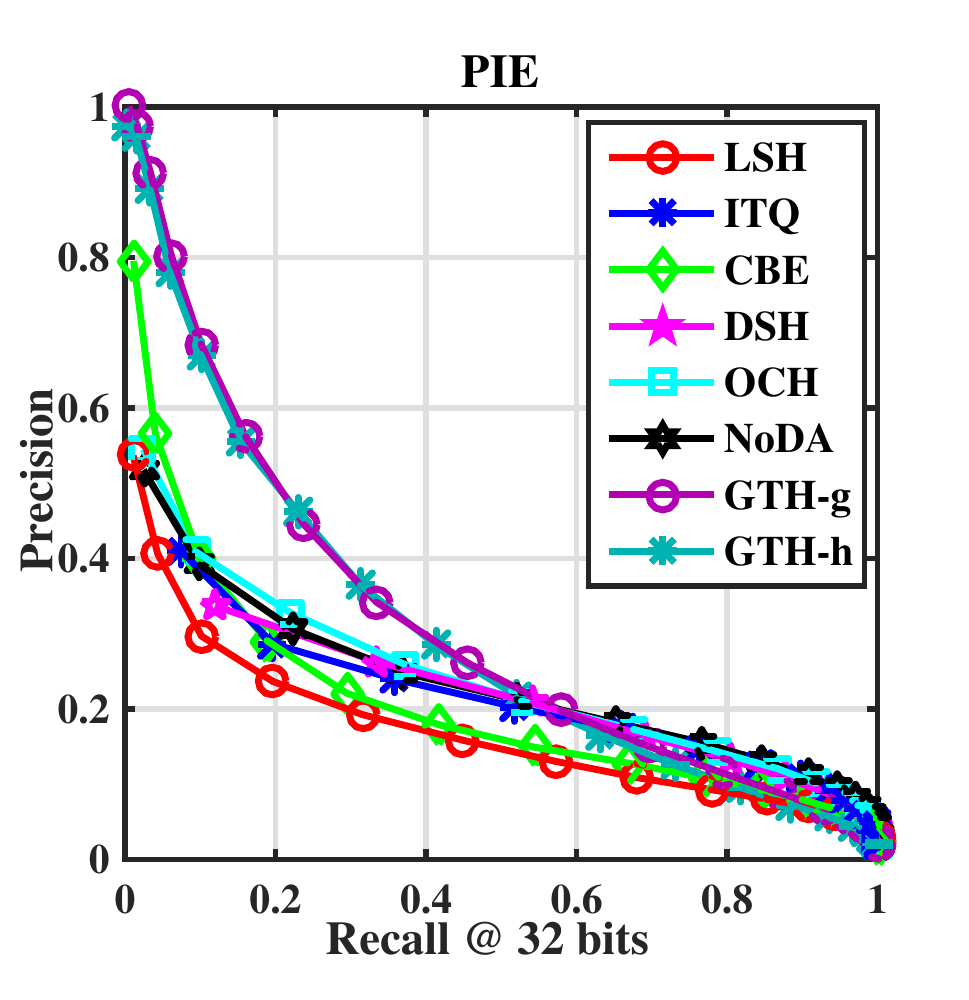}}
 \centerline{(a)}
\end{minipage}%
\hfill
\begin{minipage}{5.5cm}
 \centerline{\includegraphics[height=6cm,width=6cm]{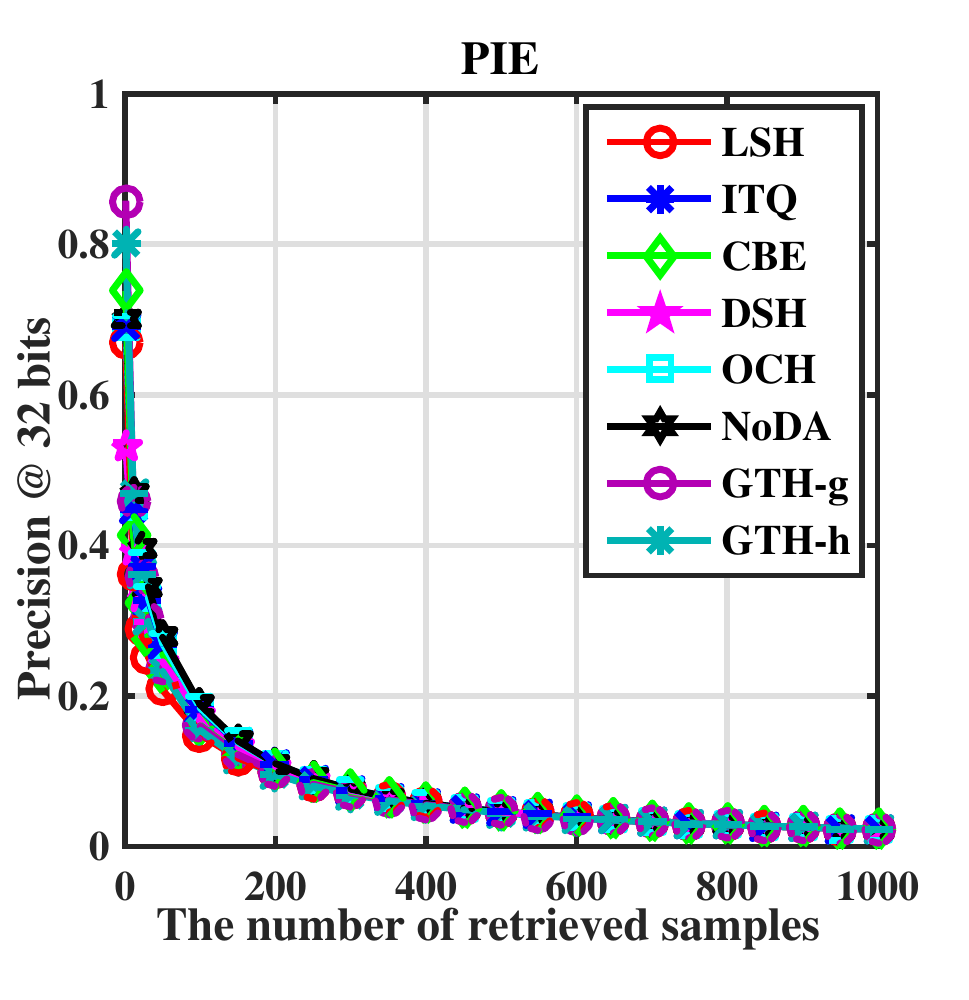}}
 \centerline{(b)}
\end{minipage}%
\hfill
\begin{minipage}{5.5cm}
 \centerline{\includegraphics[height=6cm,width=6cm]{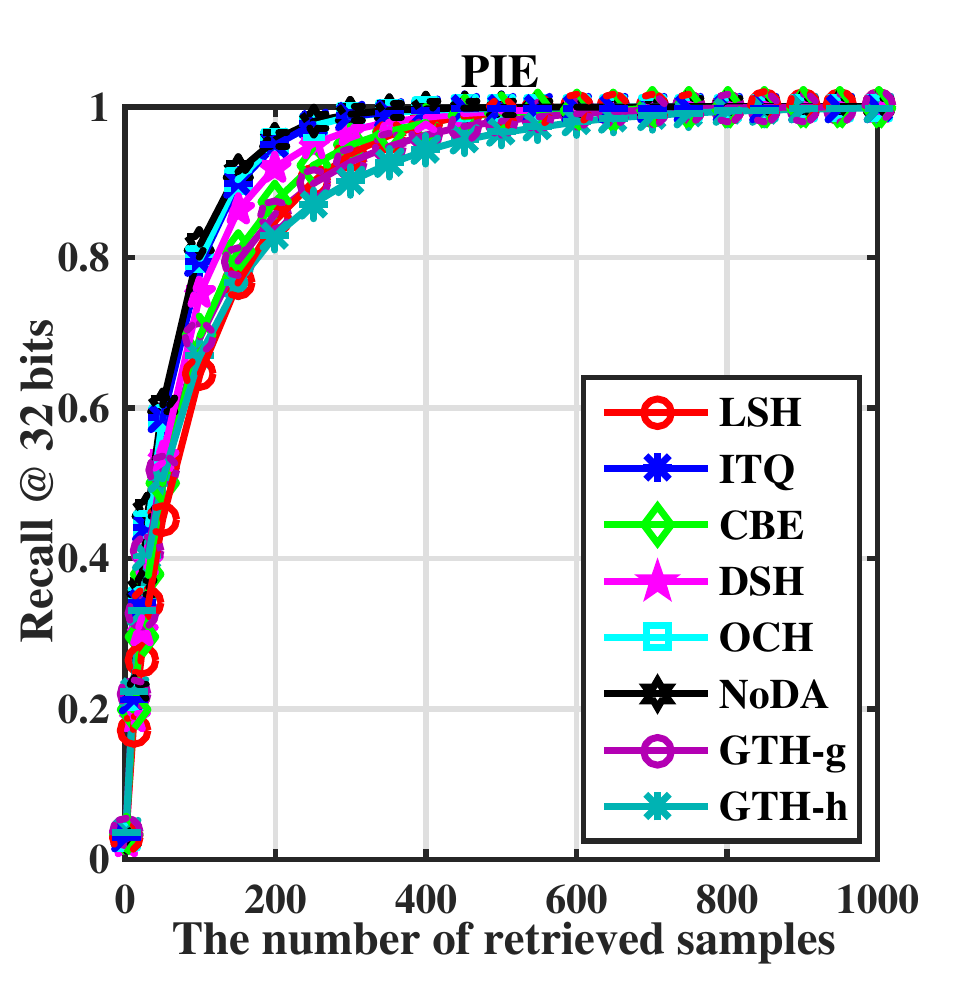}}
 \centerline{(c)}
\end{minipage}%
\caption{Retrieval performance on PIE-C29\&PIE-C05 datasets @32 bit. (a) Precision and Recall curve; (b) Precision; (c) Recall.}
\label{fig2}
\end{figure*}

\begin{figure*}
\centering
\begin{minipage}{5.5cm}
 \centerline{\includegraphics[height=6cm,width=6cm]{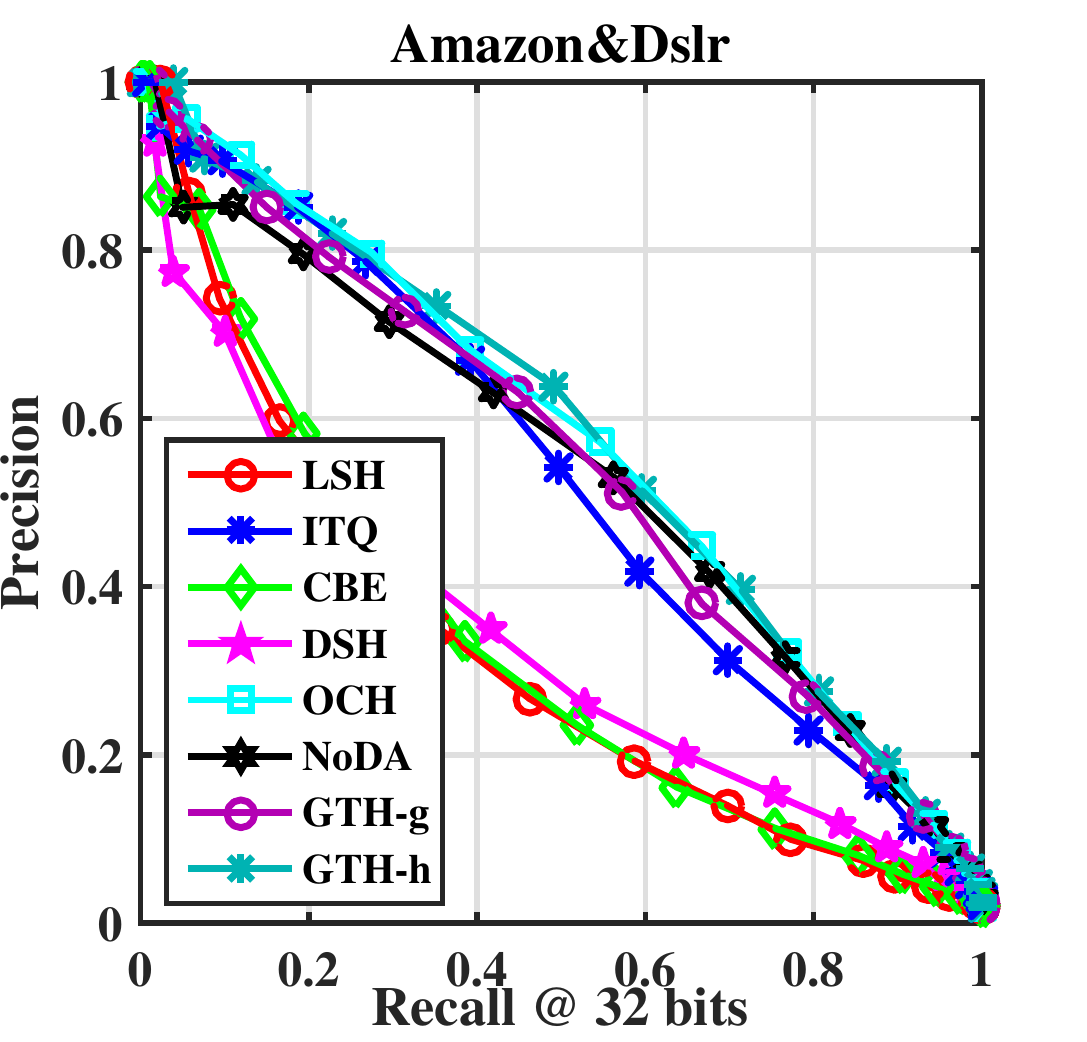}}
 \centerline{(a)}
\end{minipage}%
\hfill
\begin{minipage}{5.5cm}
 \centerline{\includegraphics[height=6cm,width=6cm]{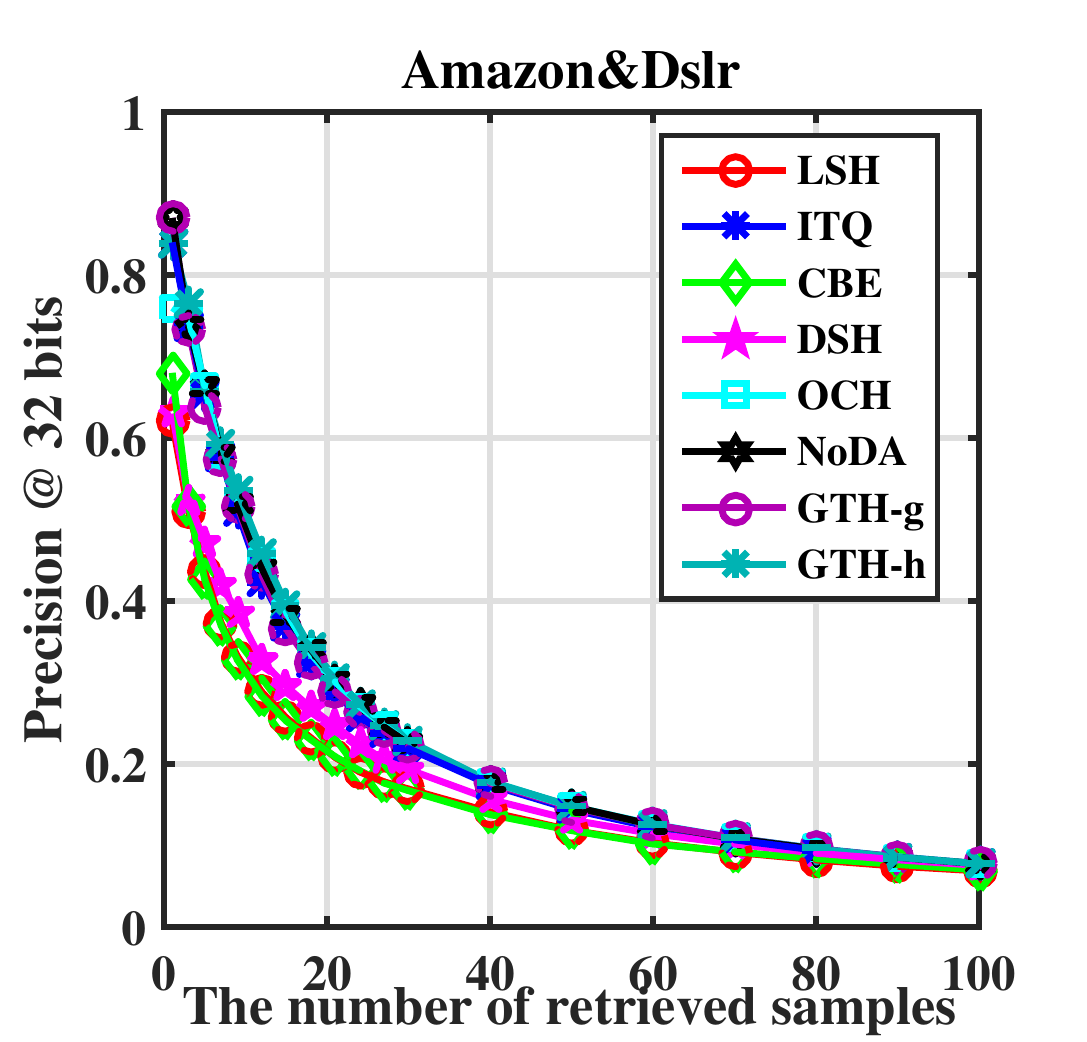}}
 \centerline{(b)}
\end{minipage}%
\hfill
\begin{minipage}{5.5cm}
 \centerline{\includegraphics[height=6cm,width=6cm]{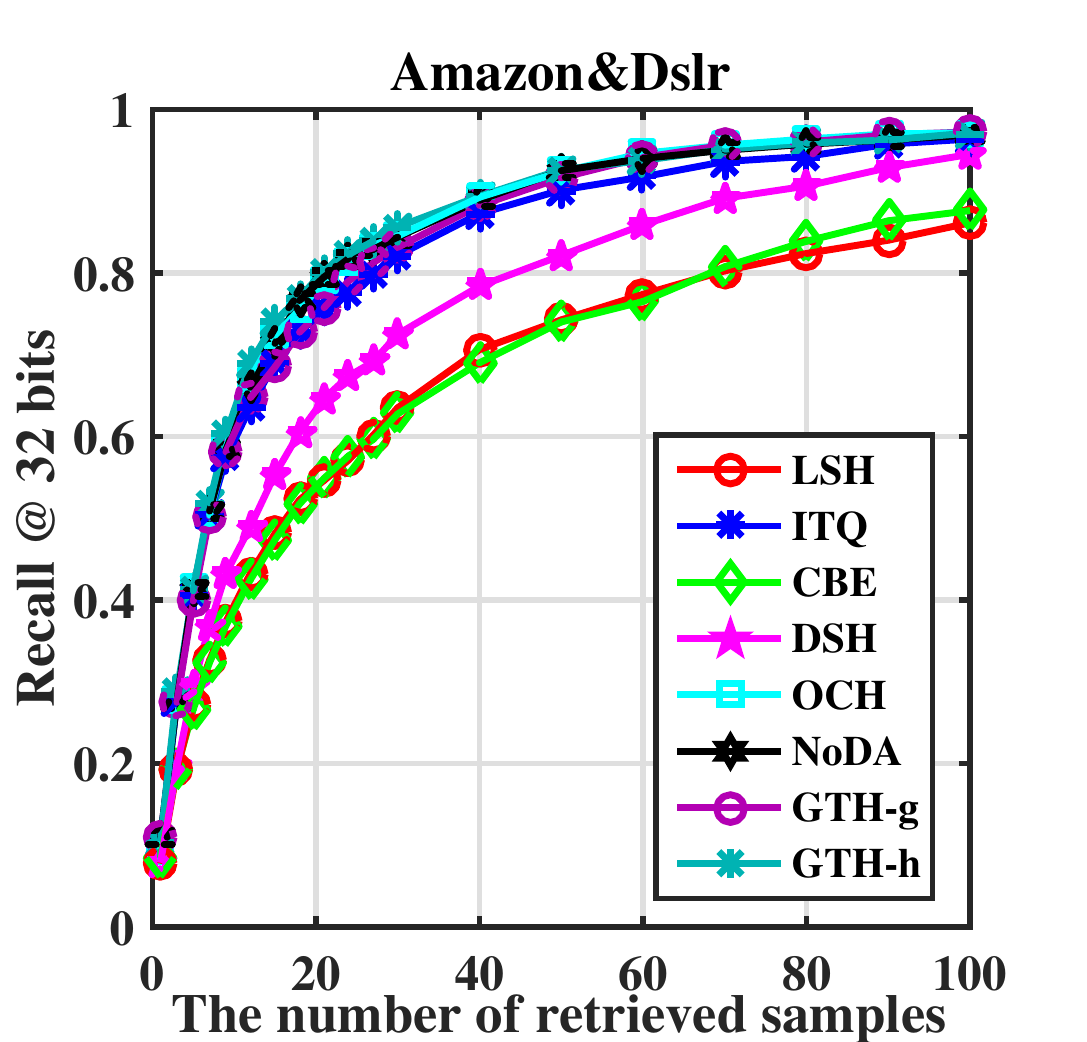}}
 \centerline{(c)}
\end{minipage}%
\caption{Retrieval performance on Amazon\&Dslr datasets @32 bit. (a) Precision and Recall curve; (b) Precision; (c) Recall.}
\label{fig3}
\end{figure*}

\begin{figure*}
\centering
\begin{minipage}{5.5cm}
 \centerline{\includegraphics[height=6cm,width=6cm]{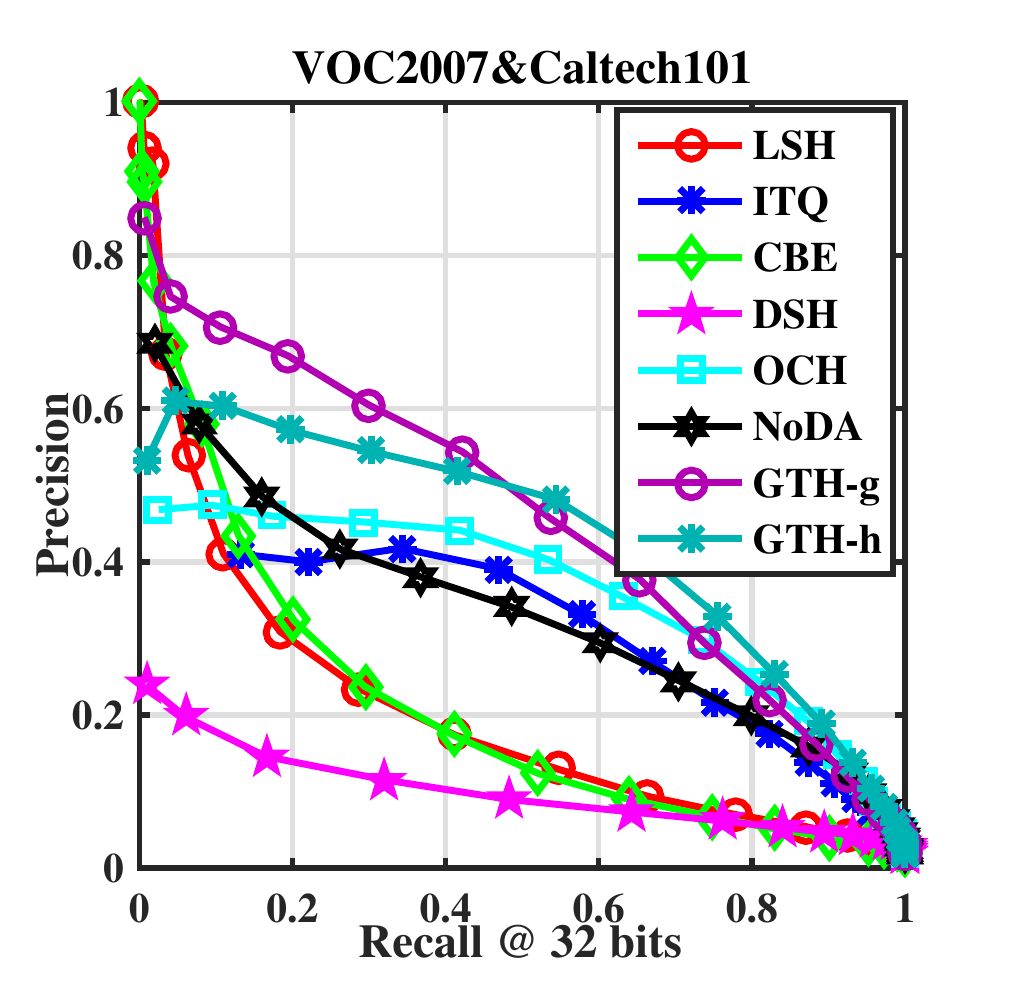}}
 \centerline{(a)}
\end{minipage}%
\hfill
\begin{minipage}{5.5cm}
 \centerline{\includegraphics[height=6cm,width=6cm]{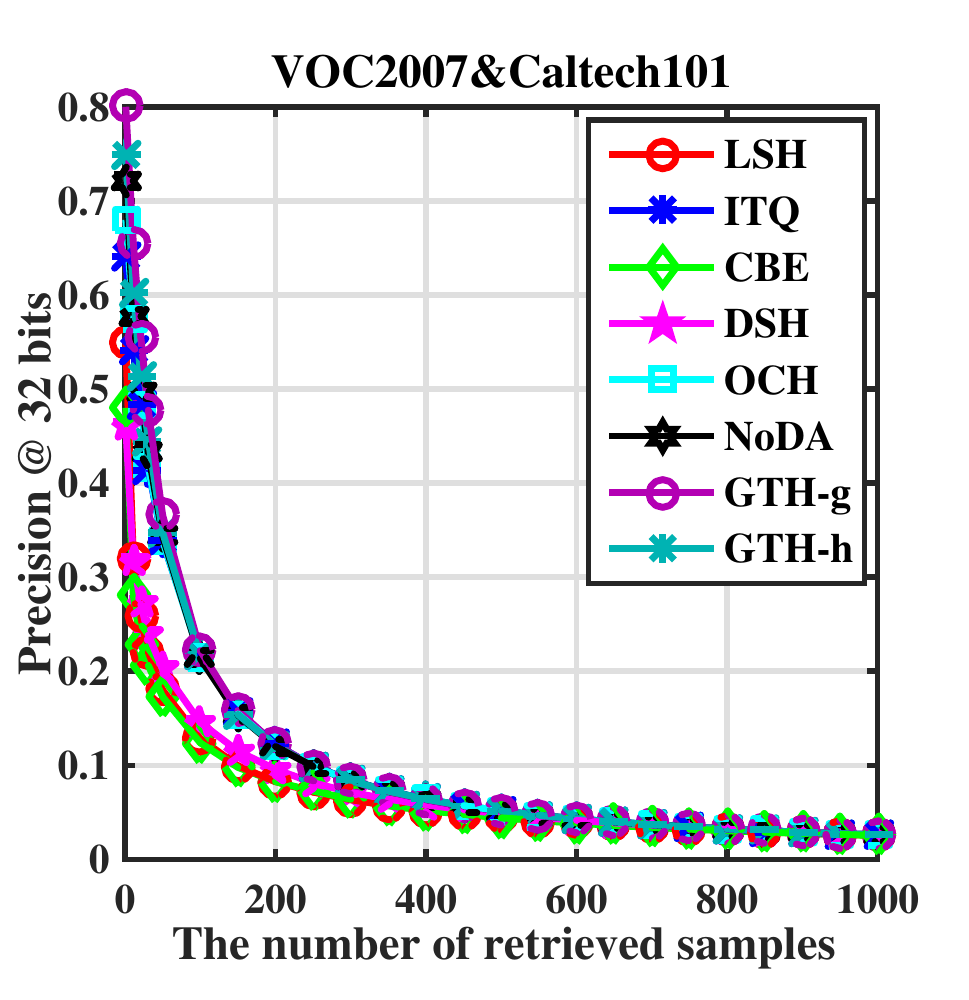}}
 \centerline{(b)}
\end{minipage}%
\hfill
\begin{minipage}{5.5cm}
 \centerline{\includegraphics[height=6cm,width=6cm]{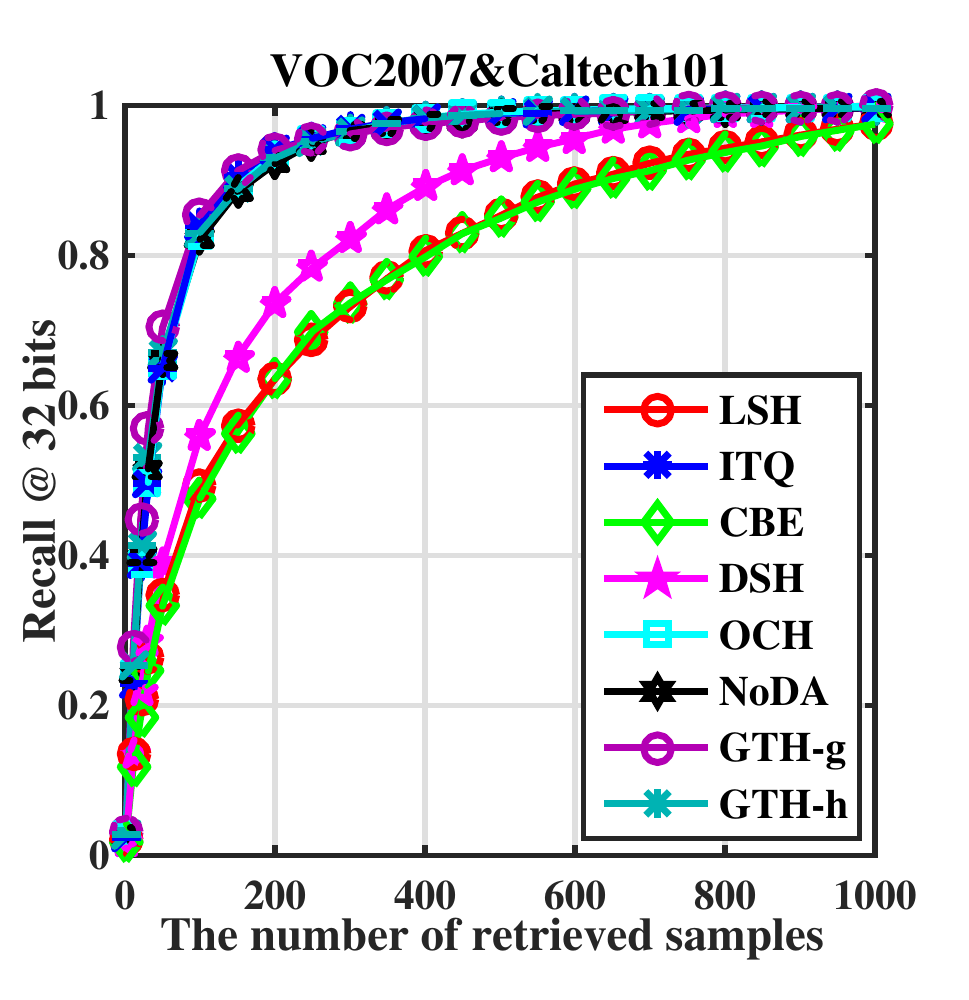}}
 \centerline{(c)}
\end{minipage}%
\caption{Retrieval performance on VOC2007\&Caltech101 datasets @32 bit. (a) Precision and Recall curve; (b) Precision; (c) Recall.}
\label{fig4}
\end{figure*}

\textbf{Parameter settings and Implementation details:} There are two trade-off parameters in the objective function (\ref{eq9}). $\lambda_1$ and $\lambda_2$ are used to penalize the loss between the binary codes and its signed magnitude. For our GTH, we empirically set $\lambda_1$ to 0.1 and $\lambda_2$ to 1.

The compared baseline methods are proposed under no domain adaption assumption. For a fair contrast, we use all the source domain data and target domain training data (except the queries on the target domain) as the model input for all compared methods. Besides, we use OCH as a NoDA method. In training phase, we use the training images in target domain as the input of NoDA method. We only focus the retrieval performance on target domain.

\textbf{Retrieval evaluation:} In the Table. \ref{tab1}, we report the MAP scores of all the compared methods and our GTH on PIE-C29\&PIE-C05, Amazon\&Dslr, and VOC2007\&Caltech101 databases. The code lengths are varying from 16 to 64. From the table, we can see that our GTH outperforms compared methods on all databases in most cases. More detailedly, our GTH-h outperforms best compared method NoDA over 4\% on PIE-C29\&PIE-C05 datasets when the code length is set as 16 bit. On Amazon\&Dslr datasets, our GTH-h outperforms best compared method OCH almost 4\% with code length set to 16. On the VOC2007\&Caltech101 databases, our GTH outperforms much more than the best compared method when the code length is set as 24, 32, 48, and 64. The above results demonstrate the effectiveness of our GTH model and our GTH is more suitable to the condition that there are not enough training images used to learn precise hashing codes in the domain of interest. We also show the PR-curve, Precision and Recall for PIE-C29\&PIE-C05 datasets as shown in Fig. \ref{fig2}, Amazon\&Dslr dataset as shown in Fig. \ref{fig3}, and VOC2007\&Caltech101 databases as shown in Fig. \ref{fig4}. The code length is set to 32 in Figures \ref{fig2}, \ref{fig3}, and \ref{fig4}. From the figures, we can see that our GTH always presents competitive retrieval performance compared to baselines, which demonstrates the efficiency of our GTH.
\begin{figure*}
\centering
\begin{minipage}{5.5cm}
 \centerline{\includegraphics[height=5.5cm,width=6cm]{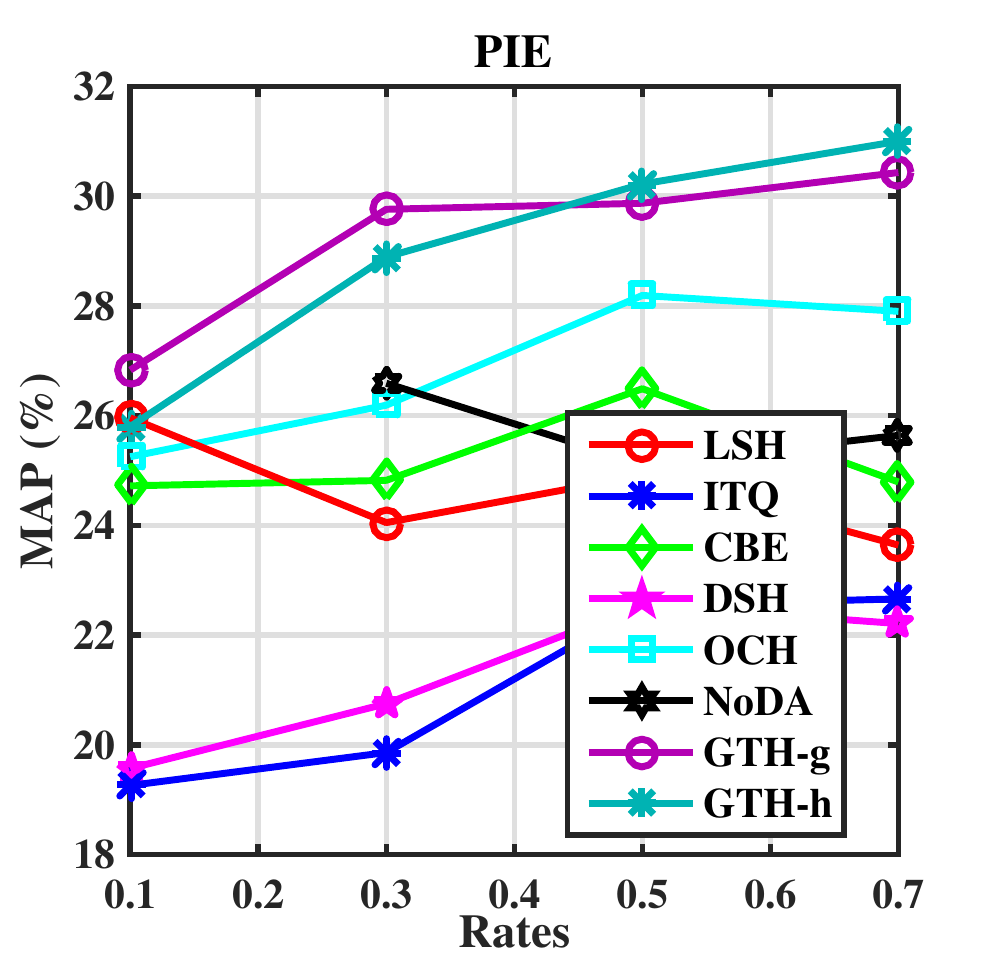}}
 \centerline{(a)}
\end{minipage}%
\hfill
\begin{minipage}{5.5cm}
 \centerline{\includegraphics[height=5.5cm,width=6cm]{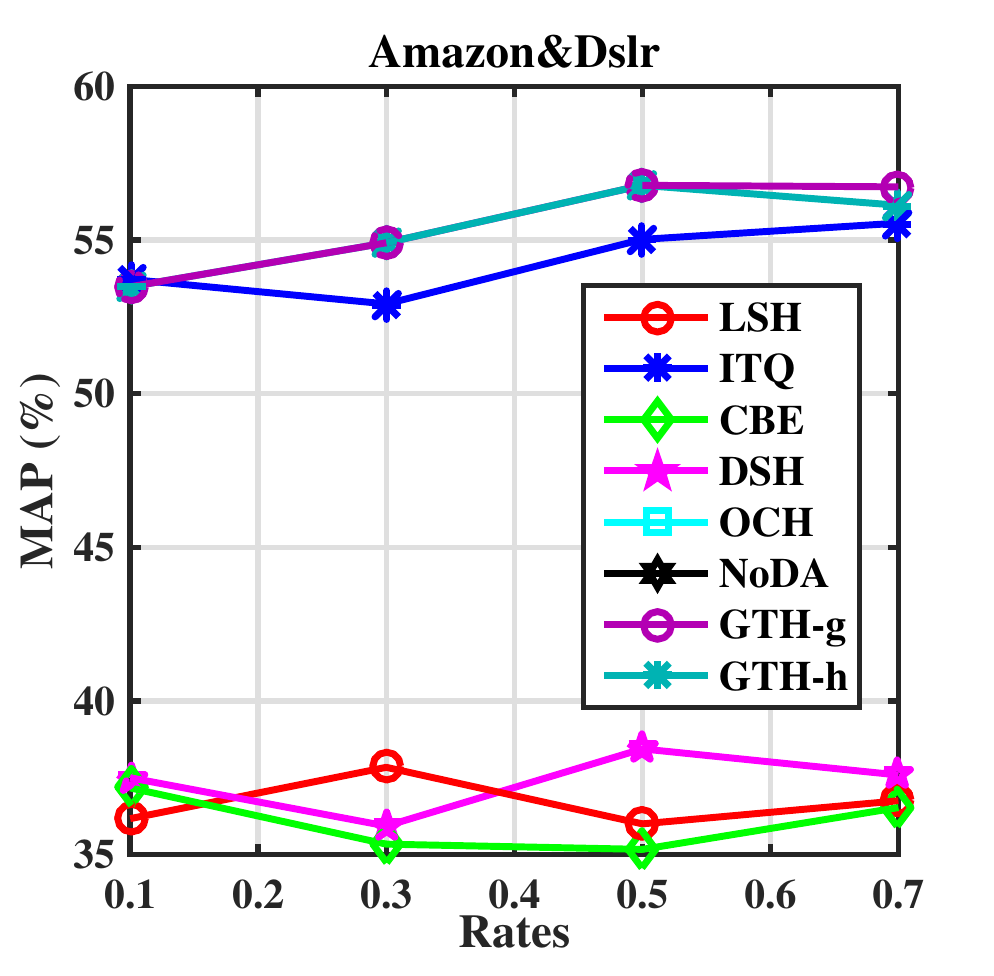}}
 \centerline{(b)}
\end{minipage}%
\hfill
\begin{minipage}{5.5cm}
 \centerline{\includegraphics[height=5.5cm,width=6cm]{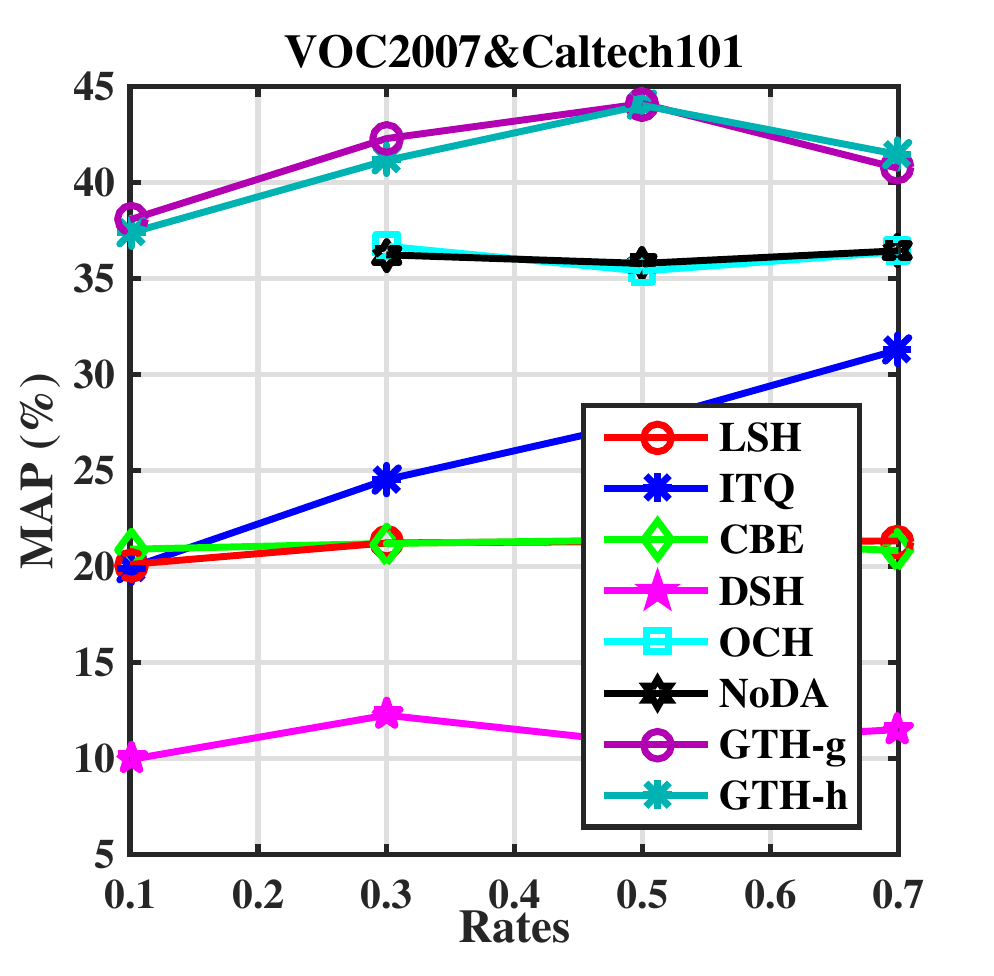}}
 \centerline{(c)}
\end{minipage}%
\caption{MAP scores @32 bit with varying number training images of target domain. (a) PIE-C29\&PIE-C05; (b) Amazon\&Dslr; (c) VOC2007\&Caltech101.}
\label{fig5}
\end{figure*}
\subsection{Retrieval evaluation on varying target training numbers}
In order to further demonstrate the efficiency of our GTH by using less target training data, we use different numbers of training data on target domain to learn the hashing functions. Specially, we choose 10\%, 30\%, 50\%, and 70\% images from training data of target domain as training data i.e., model input. After training, we also use testing hashing codes to search the most similar hashing codes in the whole training samples. The experiments are conducted on PIE-C29\&PIE-C05, Amazon\&Dslr, and VOC2007\&Caltech101 databases respectively. The MAP scores of all compared methods and our GTH are shown in Fig. \ref{fig5}. Due to the input number limitation of OCH method, there are empty MAP scores in some cases. The code length is set as 32. It is worth noting that our GTH always outperforms all the compared methods, which further demonstrates the efficiency of our GTH on the condition that there are less target domain samples to learn precise hashing codes on the domain of interest.
\subsection{Parameters Sensitivity}

\begin{figure}
\centering
\begin{minipage}{4.2cm}
 \centerline{\includegraphics[height=4cm,width=4.2cm]{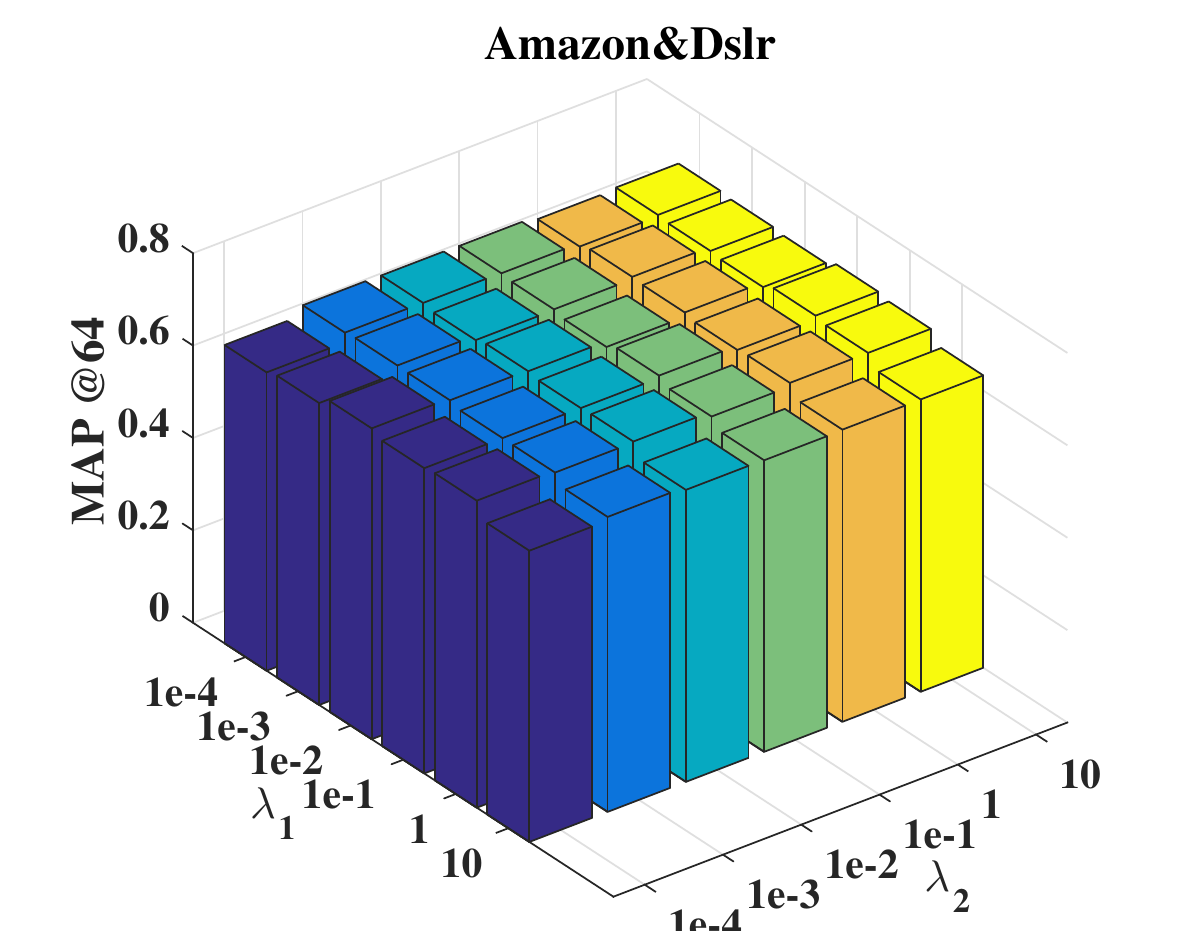}}
 \centerline{(a)}
\end{minipage}%
\hfill
\begin{minipage}{4.2cm}
 \centerline{\includegraphics[height=4cm,width=4.2cm]{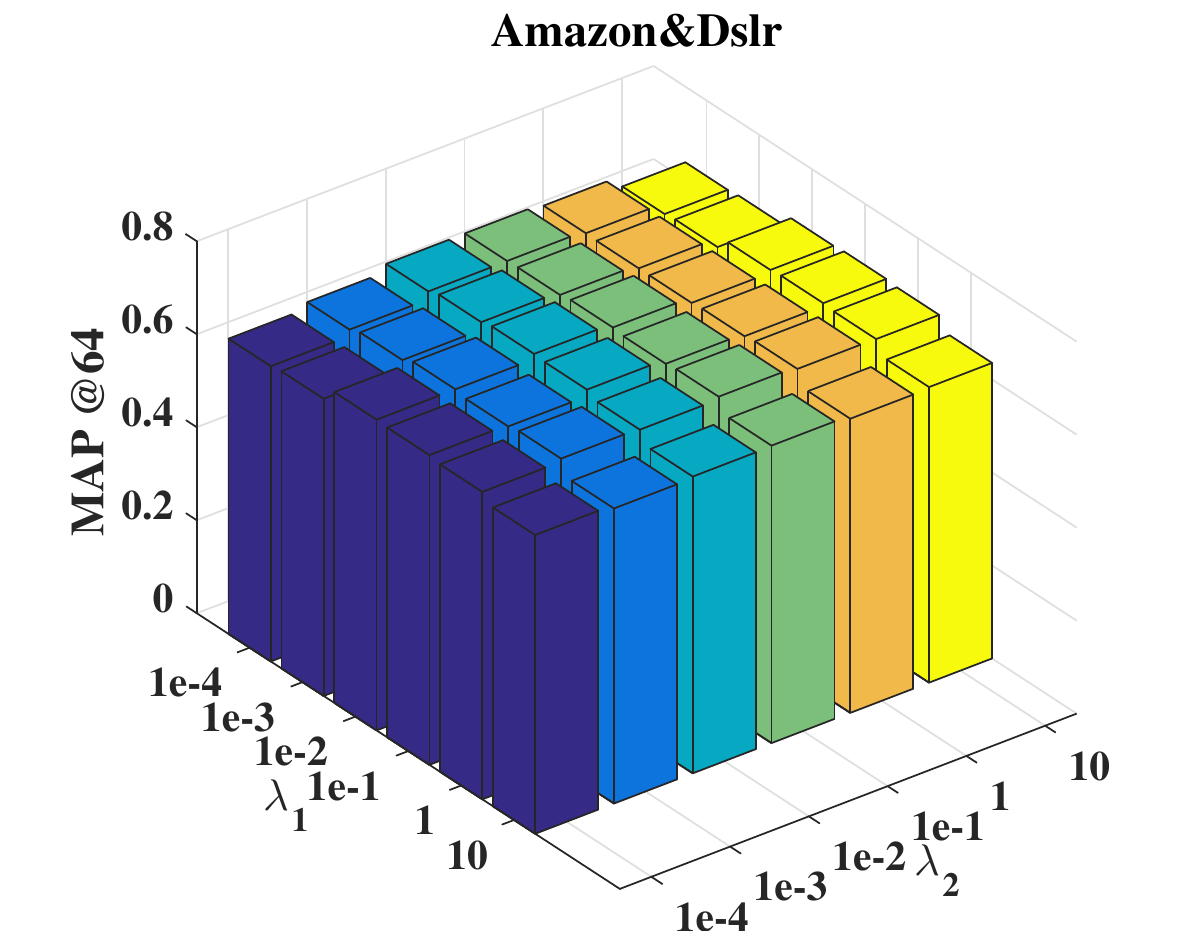}}
 \centerline{(b)}
\end{minipage}%
\caption{Parameters Sensitivity. (a) GTH-g; (b) GTH-h.}
\label{fig6}
\end{figure}
In order to further investigate the properties of the proposed method, the retrieval performances versus the different values of regularization parameters, $\lambda_1$ and $\lambda_2$, are explicitly explored. To clearly show the results, we perform experiments on  Amazon\&Dslr databases to verify the parameters sensitivity. Specifically, we tune the value of both parameters from \{0.0001, 0.001, 0.01, 0.1, 1, 10\}. The MAP scores with code length set to 64 are shown in Fig. \ref{fig6}. We can observe that the performances of our GTH-g and GTH-h models are not very sensitive to the settings of $\lambda_1$ and $\lambda_2$. Apparently, when the parameters are not very large, the MAP scores of our methods are not severely influenced. This also demonstrates that both regularization terms are indispensable for superior performances. Overall, the proposed models are not sensitive to the parameters in a reasonable range.

\section{Conclusion}
We propose a simple but effective transfer hashing method named Optimal Projection Guided Transfer Hashing (GTH) in this paper. Inspired by transfer learning, we propose to borrow the knowledge from a related but different domain. We assume that similar images between target and source domains should mean small discrepancy between hashing projections. Therefore, we let the projections of target and source domain close to each other so that the similar instances between those two domain will be transformed into similar hashing codes. We propose the GTH model from the view of maximum likelihood estimation in this paper and design a iteratively weighted $l_2$ loss for the errors between the projections of source and target domains, which makes our GTH more adaptive to cross-domain case. Extensive experiments on three groups benchmark databases have been conducted. The experimental results show that our GTH always show much higher retrieval performance when there are much less target samples, which verify that our method outperforms many state-of-the-art learning to hash methods.
\section{Acknowledgement} This work is supported by National Natural Science Fund of China (Grant 61771079) and the Fundamental Research Funds for the Central Universities.
\bibliography{egbib}
\bibliographystyle{aaai}
\end{document}